\documentclass[superscriptaddress,secnumarabic,amssymb,amsmath,nobibnotes,aps,prd,nofootinbib,preprint]{revtex4}

\usepackage{booktabs}
\usepackage{mathrsfs}
\usepackage{graphicx}
\usepackage{latexsym}
\usepackage{amsmath}
\usepackage{amsfonts}
\usepackage[dvips]{epsfig}
\usepackage{color}

\begin{document}

\title{Quantum extension of the noncommutative Yukawa potential: spectrum, thermodynamics, and short-distance collapse instability}

\author{Albert C. R. Mendes}\email{albertcrm@ufjf.br}
\affiliation{Department of Physics, Universidade Federal de Juiz de Fora, MG, Brazil}

\author{M. J. Neves}\email{mariojr@ufrrj.br}
\affiliation{Department of Physics, Universidade Federal Rural do Rio de Janeiro, RJ, Brazil}

\begin{abstract}

The semiclassical treatment for the Yukawa
potential in noncommutative (NC) phase space developed by the ref. \cite{Sousa2026} is extended to the quantum regime in this paper. From the radial Hamiltonian obtained in this reference, we quantize the radial sector with the prescription of the conserved angular momentum of $M\to\hbar \, \sqrt{\ell(\ell+1)}$. Within this reduced model, the resulting
spectrum preserves the $(2\ell+1)$-fold $m_\ell$ degeneracy. We derive a
closed-form first-order NC energy correction in which the screening
dependence is retained exactly in the perturbation matrix element, while the zeroth-order reference spectrum remains Coulombian. We then distinguish the classical loss of a circular-orbit minimum from the short-distance quantum fall to the center behavior generated by the attractive $1/r^3$ NC term.
For the cutoff-regularized problem, direct diagonalization of the full Hamiltonian reveals a threshold $\Theta_c^{\rm qu}(\ell)$ for the appearance of a state localized near the inner cutoff $(b)$. This threshold depends on the cutoff and decreases in the sense that the underlying singular behavior is
exposed as $b\to0$. The thermodynamic quantities are evaluated from the full
numerical spectrum, and are validated
against the confined-hydrogen limit. In the parameter range studied,
first-order perturbation theory overestimates the NC correction to the
partition function by about a factor of four at a representative point,
whereas the exact finite-system heat capacity remains non-negative and its
Schottky structure is only weakly modified. Finally, a simplified
proton-neutron application gives $\Theta_c^{\rm qu}(\ell=1)\approx1$ fm$^2$
for a hard-core radius $b=0.4$ fm. Because this scale depends on the
short-distance model, it is interpreted as a model-dependent consistency
scale rather than as a precision constraint.
\end{abstract}
\maketitle

\section{Introduction}

Noncommutative phase-space quantum mechanics was introduced as an effective
description of possible low-energy corrections associated with a nontrivial
space-time structure at the Planck scale; in few-body systems it shifts the
energy levels of familiar central potentials, so that the available
spectroscopic precision can be converted into bounds on the
noncommutativity parameter \cite{GamboaLoeweRojas2001,Chaichian2001}. The
thermodynamic properties of non-relativistic particles subject to central
potentials form, in turn, a well established line of study, supported by
approximate analytical methods and with applications to molecular
spectroscopy \cite{Okorie2018,Dong2007}; in this context, confinement and
the discrete level structure produce characteristic features in the
thermodynamic functions, such as Schottky anomalies, while negative heat
capacities indicate long-range interactions or nonadditivity
\cite{LyndenBell1999,Campa2009}. These two lines meet in a natural
question: whether a noncommutative deformation of phase space changes the
thermodynamics of few-body quantum systems in a consistent way, and under
which conditions such changes remain well defined.

Motivated by these considerations and from ref. \cite{Sousa2026}, the authors established a semiclassical treatment of the Yukawa and Lee--Wick
potentials in a phase space deformed by the noncommutativity spacetime, obtaining the
density of states, partition function, mean energy, and heat capacity in the microcanonical and canonical ensembles. More generally, phase-space noncommutativity and its quantum-mechanical realizations have
been formulated in several equivalent deformation schemes \cite{DjemaiSmail2003,BertolamiLeal2015,Kokado2003,Alavi,Maireche2020}. A central result of that work
was the identification of negative heat-capacity regions for the NC
Lee--Wick potential, interpreted as a symptom of the limits of the
semiclassical and perturbative treatment rather than a definitive physical
effect.

The present work extends that treatment to the non-relativistic quantum regime for the Yukawa potential. Rather than fully requantizing the original noncommutative phase-space algebra, we start from the radial Hamiltonian already reduced classically in \cite{Sousa2026} and quantize it directly, a choice that preserves the structure of the classical formalism and allows direct comparison between the two treatments.

The noncommutative term of the Yukawa potential, which diverges as $r^{-3}$ at the origin without shielding factor, produces a quantum \emph{center-collapse} instability, with a cutoff-dependent threshold $\Theta_c^{\rm qu}(\ell)$ that is conceptually distinct from the qualitative perturbative condition used in the ref. \cite{Sousa2026}. This threshold determined here, numerically defines the domain within which the quantum thermodynamic results presented are quantitatively reliable.

This paper is organized as follows : The section 2 establishes the radial reduction and the quantization prescription. The section 3 obtains the perturbative spectrum with exact $\kappa$-dependence in the matrix element, and discusses the resulting degeneracy structure. In the section, we extend the classical orbital stability analysis to the NC case, obtaining a closed-form for the critical parameter $\Theta_c(\ell,\kappa)$. The section 5 identifies and quantitatively characterizes the center-collapse instability. The section 6 builds the quantum partition function, mean energy, and heat capacity, with numerical validation and comparison with the classical result. In the section 7, we show a phenomenological proton-neutron scale. The section 8 concludes and points to natural extensions.

\section{Radial reduction and quantization prescription}

In the ref. \cite{Sousa2026}, the noncommutative Poisson algebra was used to eliminate the angular variable $\dot\phi$ in favor of the conserved quantity $M$, reducing the two-body problem to an effective one-particle radial Hamiltonian,
\begin{equation}
H=\frac{\mu\,\dot r^{\,2}}{2}+\frac{M^{2}}{2\mu r^{2}}+V_{NC}(r)
\quad , \quad
V_{NCYP}(r)=-\frac{k\,e^{-\kappa r}}{r}-\frac{\Theta k M\,e^{-\kappa r}}{r^{3}} \; ,
\label{eq:HclassNC}
\end{equation}
where $\mu$ is the reduced mass, $k$ the Yukawa coupling, $\kappa$ the
screening parameter, and $\Theta$ is the NC parameter. In \cite{Sousa2026}, $M$ is treated as a fixed parameter characterizing the dynamical configuration of the system.

We quantize the already-reduced radial Hamiltonian Eq.\eqref{eq:HclassNC}
directly. Promoting $p_r\to-i\hbar\partial_r$, the term $M^2/(2\mu r^2)$
must reproduce the usual centrifugal barrier in the commutative limit, which fixes $M^2 \to \hbar^2 \, \ell\,(\ell+1)$. For the term linear in $\Theta$, we
adopt the consistent prescription
\begin{equation}
M\;\longrightarrow\;\hbar \, \sqrt{\ell(\ell+1)} \; ,
\label{eq:Mprescription}
\end{equation}
in everywhere $M$ appears. This choice is adopted throughout the present reduced model. It preserves the $(2\ell+1)$-fold $m_\ell$ degeneracy: the resulting energy correction depends only on $(n,\ell)$, never on $m_\ell$, so the $2\ell+1$ sublevels of a given $\ell$ remain degenerate to any order in $\Theta$ within this scheme.

The resulting radial Schr\"odinger equation is
\begin{equation}
-\frac{\hbar^{2}}{2\mu}\,u_{n\ell}''(r)+\left[ \, \frac{\hbar^{2}\,\ell(\ell+1)}{2\mu r^{2}}
-\frac{k\,e^{-\kappa r}}{r}-\frac{\Theta \, k \,\hbar\,\sqrt{\ell(\ell+1)}\,e^{-\kappa r}}{r^{3}} \,
\right]u_{n\ell}(r)=E\,u_{n\ell}(r) \; ,
\label{eq:radial}
\end{equation}
where $u_{n\ell}(r)=r\,R_{n\ell}(r)$, with $u(b)=0$ and $u(R)=0$. Both conditions are carried over from \cite{Sousa2026} and belong to the physical hypothesis of the model, not to a technical convenience: $b$ is the hard-sphere radius, encoding the
finite size of the interacting particles and excising the short-distance
region where the NC term ($\sim r^{-3}$) would otherwise leave the Hamiltonian
unbounded from below (Section 5); $R$ is the box radius, providing the finite
thermodynamic volume and discretizing the scattering continuum, without which
the statistical sums diverge (Section 6.3) and, at the screening used
here, sustaining the $\ell\geq1$ states, which have no genuine potential well
(Section 4). 

\section{Perturbative spectrum}

\subsection{Perturbation theory and exact radial integral}

The hamiltionian operator is written as $H=H_0+\Theta \, H_1$ with
\begin{equation}
H_0=\frac{p_r^{2}}{2\mu}+\frac{\hbar^{2}\ell(\ell+1)}{2\mu r^{2}}-\frac{k\,e^{-\kappa r}}{r} 
\quad , \quad
H_1=-\frac{k\hbar\sqrt{\ell(\ell+1)}\,e^{-\kappa r}}{r^{3}} \; ,
\end{equation}
and the first-order Rayleigh--Schr\"odinger correction is
\begin{equation}
\Delta E^{(1)}_{n\ell}(\Theta,\kappa)=-\Theta \,  k \, \hbar \, \sqrt{\ell(\ell+1)}\;I(n,\ell;\kappa)
\quad , \quad
I(n,\ell;\kappa)\equiv\left\langle\frac{e^{-\kappa r}}{r^{3}}\right\rangle_{n\ell} \; .
\label{eq:dE1general}
\end{equation}
The Hamiltonian $H_0$ (pure Yukawa) has no closed-form analytic solution. We use as the zeroth-order basis the exact hydrogenic eigenfunctions (pure Coulomb potential $-k/r$), with effective Bohr radius and energy, respectively,
\begin{equation}
a_0=\frac{\hbar^{2}}{\mu \, k} 
\quad \mbox{and} \quad 
E^{(0)}_n=-\frac{\mu \, k^{2}}{2\hbar^{2}\,n^{2}} \; .
\label{eq:E0}
\end{equation}
The $\kappa$-dependence of the matrix element $I(n,\ell;\kappa)$ is retained exactly, without
series expansion: using $R_{n\ell}(r)=N_{n\ell}\,e^{-x/2}x^{\ell}
L_{n-\ell-1}^{2\ell+1}(x)$, $x=2r/(na_0)$, where $N_{n\ell} = \left( 2/n a_{0} \right)^{3/2} \sqrt{(n - \ell - 1)!/(2n \left[(n + \ell)!\right])}$ is the normalization factor of the radial wavefunction of the hydrogen atom, and since the associated Laguerre
polynomial has finite degree, the integral reduces to a finite sum of elementary gamma-function integrals :
\begin{equation}
I(n,\ell;\kappa)=N_{n\ell}^{2}\sum_{j=0}^{2(n-\ell-1)}c_j^{(n\ell)}\,
\frac{\Gamma(2\ell+j)}{s^{2\ell+j}}
\quad , \quad
s\equiv1+\frac{\kappa n a_0}{2} \; .
\label{eq:Igeneral}
\end{equation}
\begin{table}[h]
\label{tab:Inl}
\begin{tabular}{|ccc|}
\toprule
\hline\hline
$(n,\ell)$ &\,& $I(n,\ell;\kappa)$ \\
\hline\hline
\midrule
$(2,1)$ &\,& $\dfrac{1}{24\,s^{2}}$ \\[8pt]
$(3,1)$ &\,& $\dfrac{8s^{2}-8s+3}{243\,s^{4}}$ \\[8pt]
$(3,2)$ &\,& $\dfrac{1}{405\,s^{4}}$ \\[8pt]
$(4,1)$ &\,& $\dfrac{10s^{4}-20s^{3}+21s^{2}-12s+3}{384\,s^{6}}$ \\[8pt]
$(4,2)$ &\,& $\dfrac{9s^{2}-12s+5}{1920\,s^{6}}$ \\[8pt]
$(4,3)$ &\,& $\dfrac{1}{2688\,s^{6}}$ 
\\
\hline\hline
\bottomrule
\end{tabular}
\centering
\caption{The exact radial integral $I(n,\ell;\kappa)$, in units of $a_0^{-3}$,
with $s=1+\kappa n a_0/2$. Verified against standard inverse-power hydrogenic expectation-value
relations in the $\kappa\to0$ limit \cite{Marxer1991}.}
\end{table}
For $\ell=0$, Eq.\eqref{eq:Igeneral} diverges logarithmically ($s$-wave
functions have $R_{n0}(0)\neq0$). With the physical cutoff $b$, this
divergence is regularized, with universal asymptotic behavior
$I(n,0;\kappa,b)\to R_{n0}(0)^2\ln(1/b)$, $R_{n0}(0)^2=4/(n^3a_0^3)$, as
$b\to0$. Since the prefactor $\sqrt{\ell(\ell+1)}$ vanishes exactly for
$\ell=0$, this correction never contributes to the energy: $\Delta
E^{(1)}_{n,0}\equiv0$.

\subsection{Full spectrum and degeneracy}

\begin{figure}[h]
\centering
\includegraphics[width=0.72\textwidth]{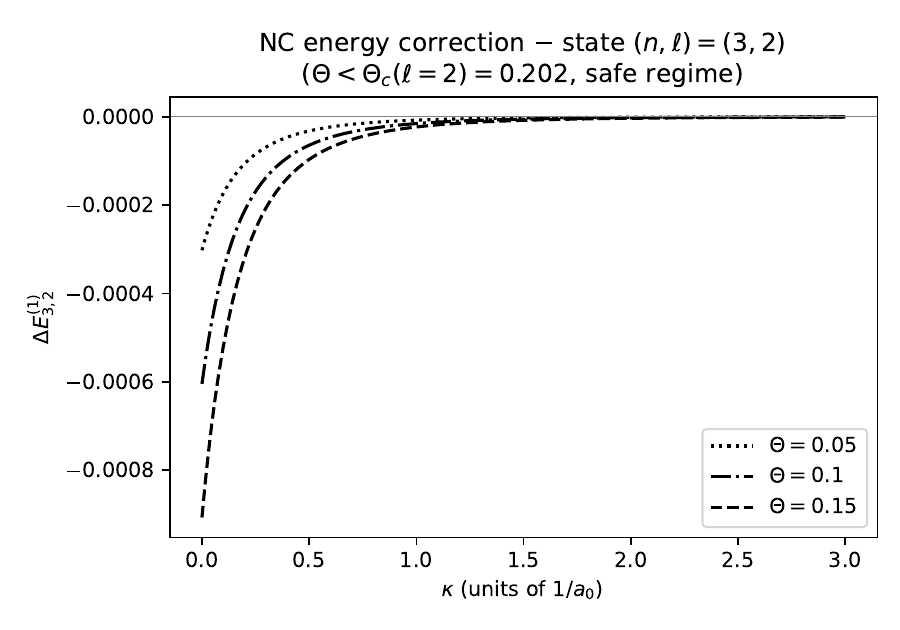}
\caption{Energy correction $\Delta E^{(1)}_{3,2}(\Theta,\kappa)$ as a function of $\kappa$, for $\Theta=0.05,0.10,0.15$.}
\label{fig:dE_vs_kappa}
\end{figure}

\begin{figure}[h]
\centering
\includegraphics[width=0.69\textwidth]{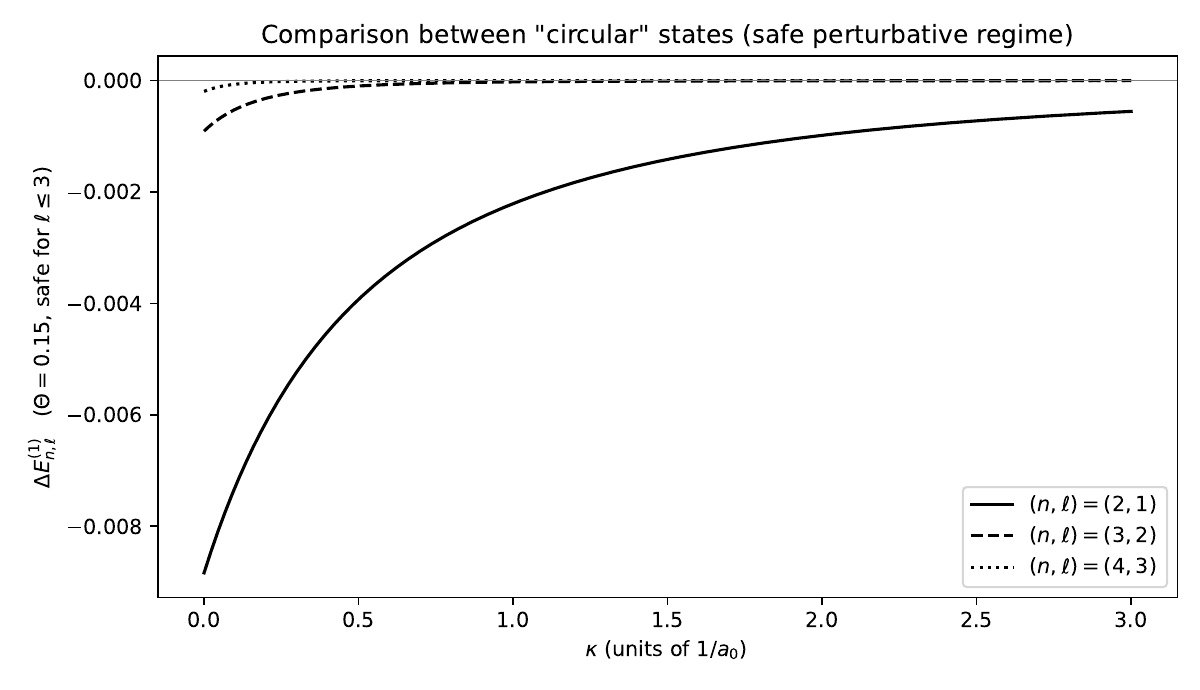}
\caption{Comparison of $\Delta E^{(1)}_{n,\ell}$ between the circular states
$(2,1)$, $(3,2)$, and $(4,3)$, at fixed $\Theta=0.15$.}
\label{fig:dE_vs_n}
\end{figure}

\begin{figure}[h]
\centering
\includegraphics[width=0.72\textwidth]{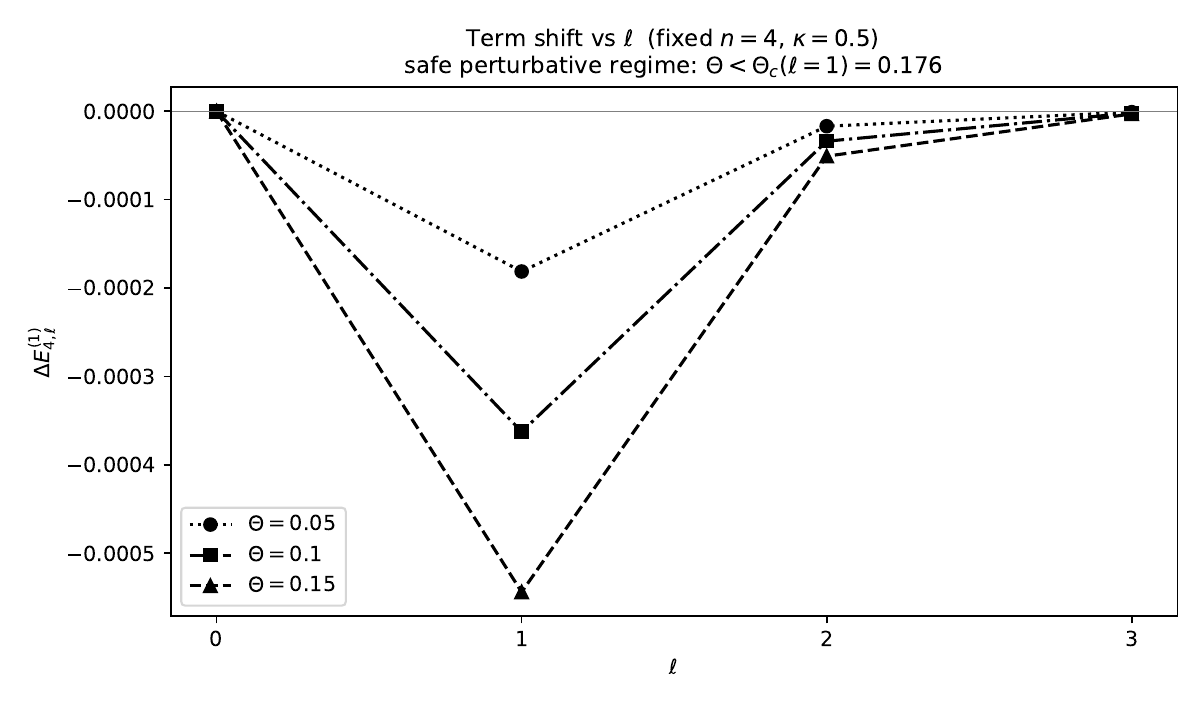}
\caption{Term shift $\Delta E^{(1)}_{4,\ell}$ as a function of $\ell$. We choose $n=4$ and $\kappa=0.5$.}
\label{fig:term_shift}
\end{figure}

Combining Eq.\eqref{eq:E0}--Eq.\eqref{eq:Igeneral}, the full spectrum in the linear $\Theta$-order is

\begin{equation}
E_{n\ell}(\Theta,\kappa)=-\frac{\mu k^{2}}{2\hbar^{2}n^{2}}
-\Theta \, k\hbar \, \sqrt{\ell(\ell+1)}\,I(n,\ell;\kappa)+\mathcal{O}(\Theta^{2}) \;
\qquad(\ell\geq1) \; ,
\label{eq:fullspectrum}
\end{equation}
with $E_{n,0}=E_n^{(0)}$, exactly. Two explicit energy levels are :
\begin{align}
E_{2,1}(\Theta,\kappa)&=-\frac{\mu k^{2}}{8\hbar^{2}}
-\frac{\sqrt2\,\Theta\,\mu^{5} \, k^{6}}{24\,\hbar^{5}\big(\hbar^{2}\,\kappa+k\mu\big)^{2}} \; , 
\\
E_{3,2}(\Theta,\kappa)&=-\frac{\mu k^{2}}{18\hbar^{2}}
-\frac{16\sqrt6\,\Theta\,\mu^{7} \, k^{8}}{405\,\hbar^{5}\big(3\hbar^{2}\,\kappa+2k\mu\big)^{4}} \; .
\end{align}
The figure \ref{fig:dE_vs_kappa} illustrates $\Delta E^{(1)}_{3,2}$ as a function of $\kappa$; Figure \ref{fig:dE_vs_n} compares the correction magnitude
between circular states of different $n$; and the figure \ref{fig:term_shift}
shows the term shift as a function of $\ell$ at fixed $n$, revealing
non-monotonic behavior: it grows in magnitude from $\ell=0$ (null) to a peak
at $\ell=1$, then decreases, reflecting the competition between the growing
prefactor $\sqrt{\ell(\ell+1)}$ and the faster-decreasing centrifugal
suppression of $I(n,\ell;\kappa)$.
Regarding degeneracy: the accidental Coulomb degeneracy between states of
equal $n$ and different $\ell$ is already broken by the screening
$\kappa\neq0$ even at $\Theta=0$; the NC correction Eq.\eqref{eq:fullspectrum}
adds to this pre-existing breaking, without depending on $m_\ell$.

\section{Classical orbital stability}

We define the effective potential, already quantized in the angular
variable ( in units of $\hbar=\mu=k=1$),
\begin{equation}
V_{\rm eff}(r;\ell,\kappa,\Theta)=\frac{\ell(\ell+1)}{2r^2}-\frac{e^{-\kappa r}}{r}
-\Theta \, \sqrt{\ell(\ell+1)}\,\frac{e^{-\kappa r}}{r^3} \; ,
\label{eq:Veff}
\end{equation}
whose critical points satisfy
\begin{equation}
V_{\rm eff}'(r)=-\frac{\ell(\ell+1)}{r^3}+\frac{e^{-\kappa r}(\kappa r+1)}{r^2}
+\Theta\sqrt{\ell(\ell+1)}\,\frac{e^{-\kappa r}(\kappa r+3)}{r^4}=0 \; .
\label{eq:dVeff}
\end{equation}
A local minimum corresponds to a stable circular orbit; its disappearance
with increasing $\Theta$ signals an NC-induced classical instability. For
$\ell=0$, $V_{\rm eff}$ does not depend on $\Theta$.

\begin{figure}[h]
\centering
\includegraphics[width=0.75\textwidth]{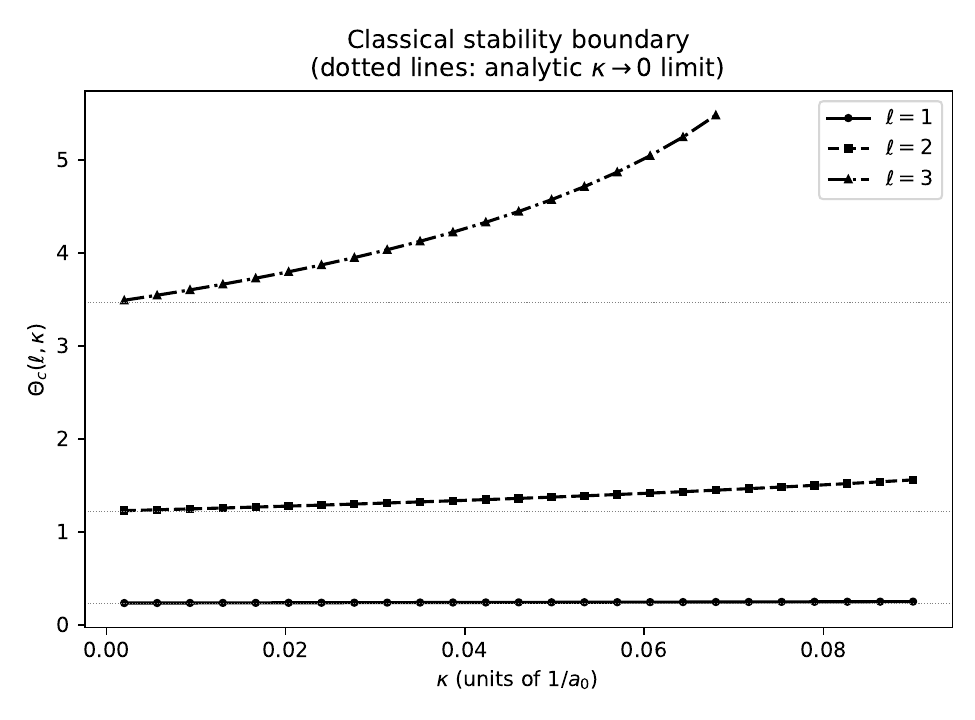}
\caption{Classical stability boundary $\Theta_c(\ell,\kappa)$ for
$\ell=1,2,3$. Below the curve a stable circular orbit exists; above, it does
not. The horizontal lines mark the analytic $\kappa\to0$ limit
Eq.\eqref{eq:ThetaC}.}
\label{fig:thetac}
\end{figure}

At $\kappa=0$ (Coulombian limit), Eq.\eqref{eq:dVeff} reduces to a quadratic equation in $r$, whose discriminant defines the critical value
\begin{equation}
\Theta_c(\ell,\kappa{=}0)=\frac{[\,\ell(\ell+1)\,]^{3/2}}{12}
\quad , \quad
r_c=\frac{\ell(\ell+1)}{2} \; .
\label{eq:ThetaC}
\end{equation}
For $\Theta>\Theta_c$, there is no longer a stable circular orbit in the Coulomb-limit effective potential. This classical statement concerns the disappearance of a local minimum; it should not be identified with the quantum fall to the center phenomenon discussed below. We numerically verified that $\Theta_c(\ell,\kappa)$ converges exactly to Eq.\eqref{eq:ThetaC} as $\kappa\to0$, and grows smoothly with
$\kappa$ (Figure \ref{fig:thetac}).
The screening value used throughout the rest
of this work ($\kappa=0.5\,a_0^{-1}$), no minimum is found for
$\ell=1,2,3$, for any $\Theta$: screening alone already eliminates the
circular orbit that would exist in the Coulomb limit. This implies that the
$\ell\geq1$ states observed in the numerical diagonalization of Section 6
are not states bound by a genuine potential well, but sustained only by the
external confinement.

\section{Fall to the center and the cutoff-dependent quantum threshold}
\label{sec:fallcenter}

The NC term of the Yukawa potential,
$-\,\Theta\sqrt{\ell(\ell+1)}\,e^{-\kappa r}/r^3$, is not suppressed as
$r\to0$. Since $1/r^3$ grows faster than the centrifugal barrier $1/r^2$, an
non regularized attractive problem is subject to the familiar fall to the center
pathology \cite{Case1950,PerelomovPopov1970}. For the finite inner radius $b$
used here, however, the Hamiltonian is a well-defined cutoff-regularized
problem. The quantity $\Theta_c^{\rm qu}$ therefore denotes the onset of a
cutoff-localized, deeply bound state, not a finite threshold below which the
non regularized Hamiltonian is mathematically stable.

This phenomenon is different from the classical instability of Section 4,
which concerns the existence of minimum for the effective potential. We
determine $\Theta_c^{\rm qu}(\ell)$ by numerically diagonalizing the full Hamiltonian, without expansion in $\Theta$, using finite differences for
$\ell=1$, $\kappa=0.5$, and $b=0.05\,a_0$:
The numerical threshold is characterized by four checks. The lowest
state remains close to the unperturbed spectrum up to $\Theta\simeq0.175$
and becomes strongly localized near the inner boundary above this value.
Refining the grid by a factor of about $40$ changes the threshold only
from $0.17511$ to $0.17569$, while varying the outer boundary has only a small effect on the localized state’s
energy. In contrast, reducing the inner cutoff produces a rapid increase in the
magnitude of the binding energy. The threshold also increases with b over the
range examined. These tests indicate that the observed instability is controlled
by the short-distance part of the model rather than by the outer numerical
boundary \cite{Plestid,LiYe}.
\begin{table}[h]
\begin{tabular}{|ccc|}
\toprule
\hline\hline
$\ell$ &\,& $\Theta_c^{\rm qu}$ \\
\hline\hline
\midrule
1 &\,& 0.176 \\
2 &\,& 0.202 \\
3 &\,& 0.229 \\
\hline\hline
\bottomrule
\end{tabular}
\centering
\caption{Cutoff-dependent quantum threshold $\Theta_c^{\rm qu}(\ell)$ at
$b=0.05\,a_0$ and $\kappa=0.5\,a_0^{-1}$.}
\label{tab:thetac_standard}
\end{table}

\begin{table}[h]
\begin{tabular}{|ccccc|}
\toprule
\hline\hline
$\Theta$ & $R=5$ & $R=10$ & $R=20$ & $R=40$ \\
\hline\hline
\midrule
0.3 & -102.26 & -102.27 & -102.29 & -102.37 \\
0.7 & -1014.90 & -1014.96 & -1015.22 & -1016.26 \\
1.5 & -4036.58 & -4036.95 & -4038.42 & -4044.28 \\
\hline\hline
\bottomrule
\end{tabular}
\centering
\caption{Energy of the cutoff-localized state versus outer box radius
($\ell=1$, $\kappa=0.5\,a_0^{-1}$, $b=0.05\,a_0$). The weak dependence on
$R$ shows that the state is controlled by the inner region.}
\label{tab:indepR}
\end{table}

\begin{table}[h]
\begin{tabular}{|ccccccccccc|}
\toprule
\hline\hline
$\Theta$ &\,& $b=0.1$ &\,& $b=0.05$ &\,& $b=0.02$ &\,& $b=0.01$ &\,& $b=0.005$ \\
\hline\hline
\midrule
0.3 &\,& 0.07 &\,& -102.27 &\,& -7583 &\,& -$1.05\times10^{5}$ &\,& -$1.22\times10^{6}$ \\
0.7 &\,& -40.23 &\,& -1014.96 &\,& -$3.32\times10^{4}$ &\,& -$3.73\times10^{5}$ &\,& -$3.83\times10^{6}$ \\
\hline\hline
\bottomrule
\end{tabular}
\centering
\caption{Energy of the cutoff-localized state as the inner cutoff is
reduced ($\ell=1$, $\kappa=0.5\,a_0^{-1}$, $R=10\,a_0$).}
\label{tab:divb}
\end{table}

For the parameters studied, $\Theta_c^{\rm qu}$ increases with $\ell$
(verified up to $\ell=7$), so $\ell=1$ sets the smallest threshold in the
range examined. The thermodynamic calculations therefore use
$\Theta=0.05,0.10,0.15$, all below the $\ell=1$ threshold. The $\ell=0$
sector is unaffected by the NC correction in the adopted prescription because
$\sqrt{\ell(\ell+1)}=0$.
\section{Quantum thermodynamics}
\subsection{Partition function, mean energy, and heat capacity}
With the $(2\ell+1)$-degeneracy and the spectrum Eq.\eqref{eq:fullspectrum}, the partition function associated with these levels energy is defined by
\begin{eqnarray}
Z_{NCYP}(\beta)&=&\sum_{n,\ell}(2\ell+1)\,e^{-\beta E_{n\ell}(\Theta,\kappa)}
\;\approx\;\sum_{n,\ell}(2\ell+1)\,e^{-\beta E_n^{(0)}}
+ 
\nonumber \\ 
&+& \Theta \, k\hbar\beta\sum_{n,\ell\geq1}(2\ell+1)\sqrt{\ell(\ell+1)}\,
I(n,\ell;\kappa)\,e^{-\beta E_n^{(0)}}+\mathcal{O}(\Theta^{2}) \; .
\label{eq:Zexpansion}
\end{eqnarray}
Unlike an alternative prescription ($M\to\hbar m_\ell$), here there is no
sum with alternating sign over $m_\ell$, and the $\mathcal{O}(\Theta)$ term survives
integrally. Defining $R(\beta)\equiv Z_1(\beta)/Z_0(\beta)$, with
\begin{equation}
Z_0(\beta)\equiv\sum_{n,\ell}(2\ell+1)\,e^{-\beta E_n^{(0)}}
\quad , \quad
Z_1(\beta)\equiv\sum_{n,\ell\geq1}(2\ell+1)\sqrt{\ell(\ell+1)}\,I(n,\ell;\kappa)\,e^{-\beta E_n^{(0)}} \; .
\label{eq:Z0Z1def}
\end{equation}
Using these definitions, the energy mean and the heat capacity 
are, respectively, given by 
\begin{align}
\langle U\rangle_{NCYP}(\beta)&=\langle U\rangle_0(\beta)-\Theta\big[R(\beta)+\beta R'(\beta)\big] \; ,
\label{eq:Ufinal}
\\
C_V^{NCYP}(\beta)&=C_V^{0}(\beta)-\Theta\beta^2\big[2R'(\beta)+\beta R''(\beta)\big] \; ,
\label{eq:CVfinal}
\end{align}
where $\langle U\rangle_0=-\partial_\beta\ln Z_0$ and $C_V^0=\beta^2\,
\partial^2_\beta\ln Z_0$ are the commutative ($\Theta=0$) values. Written explicitly in terms of $Z_0,Z_1$ and their $\beta$-derivatives (primes), we obtain
\begin{equation}
\langle U\rangle_{NCYP}(\beta)=-\frac{Z_0'}{Z_0}
-\frac{\Theta}{Z_0^2}\Big[ \, Z_0 \, Z_1+\beta\big(Z_0 \, Z_1'-Z_1 \, Z_0'\big) \, \Big] \; ,
\label{eq:Uexplicit}
\end{equation}
and
\begin{eqnarray}
C_V^{NCYP}(\beta)&=&\frac{\beta^2}{Z_0^3}\Big[ \, Z_0^2 \, Z_0''-Z_0 \, (Z_0')^2 \, \Big] 
+ 
\\ 
&+&\frac{\Theta\beta^2}{Z_0^3}\Big[ \, -\beta Z_0^2\,Z_1''
+2Z_0(\beta Z_0'-Z_0)\,Z_1'
+\big(\beta(Z_0Z_0''-2(Z_0')^2)+2Z_0Z_0'\big)\,Z_1\Big] \; .
\nonumber
\label{eq:CVexplicit}
\end{eqnarray}

Equations (16)--(19) follow from differentiating $\ln Z$ once and twice in
$\beta$ after expanding it to first order in $\Theta$. Performing the same
differentiations without any expansion leads instead to the exact canonical
relation quoted as Eq. (20): from $\langle E\rangle = -\partial_\beta \ln Z$,
one further derivative gives
$\partial_\beta\langle E\rangle = -(\langle E^2\rangle - \langle E\rangle^2)$,
and since $\beta = 1/T$ implies $d\beta/dT = -\beta^2$, the chain rule yields
\begin{equation}
C_V(\beta,\Theta) = \frac{\partial\langle E\rangle}{\partial T}
= \beta^2\big(\langle E^2\rangle - \langle E\rangle^2\big) \geq 0 \; ,
\label{eq:CVvariance}
\end{equation}
where the moments are taken with the Boltzmann weights of the full
cutoff-regularized spectrum obtained by direct diagonalization (Sec. 6.2),
$\langle E^k\rangle = Z^{-1}\sum_{n,\ell}(2\ell+1)\,E_{n\ell}^k(\Theta,\kappa)\,
e^{-\beta E_{n\ell}(\Theta,\kappa)}$. The inequality holds because the
right-hand side is $\beta^2$ times the variance of the energy in the
Boltzmann distribution, non-negative for any normalizable weight; it is
independent of the potential and of $\Theta$. Two uses are made of it below:
the exact finite-system heat capacity cannot be negative, so any negative
region produced by the truncated formulas (16)--(19) is an artifact of the
truncation rather than a property of the spectrum; and evaluating
Eq. (20) directly on the diagonalized levels provides an independent check
of the numerical $\beta$-derivatives used in the figures.

\subsection{Non-perturbative treatment: direct diagonalization}
\label{sec:nonpert}

For the numerical thermodynamics, we diagonalize the full
Hamiltonian $H_0+\Theta H_1$ (Eq. \ref{eq:radial}) directly for each desired value of $\Theta$, obtaining the numerical spectrum $E_{n\ell}(\Theta,\kappa)$ of the full cutoff-regularized Hamiltonian, without truncated series, and building the sum
\begin{equation}
Z_{NCYP}(\beta,\Theta)=\sum_{n,\ell}(2\ell+1)\,e^{-\beta E_{n\ell}(\Theta,\kappa)} \; ,
\label{eq:Zexact}
\end{equation}
directly, without expanding the Boltzmann factor. Higher order spectral effects are therefore included in the numerical
thermodynamic quantities with no need to calculate the individual perturbative terms at higher order.

\textbf{Comparison with first order} : For the values of $\Theta$ considered here, the numerical levels differ from the first order prediction (Eq. \ref{eq:dE1general}) by up to $\sim2\%$,with the largest differences occurring for $\ell = 1$. The difference is more visible in the thermodynamic quantities: at $\beta = 20$ and $\theta = 0.15$, the numerical correction to $Z$ is about $0.098\%$, whereas the first order result gives about $0.407\%$. Thus, at this representative point, the first-order expression overestimates the numerical correction by roughly a factor of four. The present calculation does not separate
the individual second-order contributions, so we do not assign this difference to a particular perturbative term.

\subsection{Numerical implementation and validation}
\label{sec:numerics}

The free hydrogen basis makes $Z_0(\beta)=\sum_n n^2 \, e^{-\beta E_n^{(0)}}$ because the Coulomb
spectrum accumulates at zero energy. For this reason, the quantitative thermodynamic results are obtained from the cutoff-regularized Hamiltonian rather
than from this perturbative partition function. We solve Eq.\eqref{eq:radial} directly,
with $u(b)=u(R)=0$ (physical confinement $r\in(b,R)$, consistent with ref.\cite{Sousa2026}, via finite differences: tridiagonal matrix, diagonalization for
$\ell=0,\dots,9$ with $40$ states each, $b=0.05\,a_0$, $R=25\,a_0$,
$N=4000$ grid points.

As a check of the numerical method, we take $\kappa$ = 0 and compare with the
confined hydrogen problem \cite{LeyKooRubinstein1979,Aquino2007}.
 At fixed $R = 40 a_0$, reducing the inner cutoff
from $b = 0.01 a_0$ to $10^{-4}a_0$ moves the ground-state energy smoothly toward the
free Coulomb value $-0.5$, as shown in  (Table \ref{tab:hydrogen}) \cite{Chakraborty,Joshi}.
\begin{table}[h]
\begin{tabular}{|ccc|}
\toprule
\hline\hline
$b\,(a_0)$ &\,& $E_{1s}$ \\
\hline\hline
\midrule
0.01   &\,& -0.48211 \\
0.005  &\,& -0.49062 \\
0.002  &\,& -0.49612 \\
0.001  &\,& -0.49803 \\
0.0005 &\,& -0.49901 \\
0.0001 &\,& -0.49980 \\
\hline\hline
\bottomrule
\end{tabular}
\centering
\caption{Confined hydrogen ground-state energy ($\kappa=0$,
$R=40\,a_0$) as the inner cutoff $b$ is refined, converging to the exact
free value $E_{1s}=-0.5$.}
\label{tab:hydrogen}
\end{table}

The same test at the working point $\kappa=0.5$ shows that $b=0.05\,a_0$ is \emph{not} a
numerically innocuous choice (Table \ref{tab:bsens}): reducing $b$ from
$0.05$ to $0.001\,a_0$ shifts $E_{1,0}$ by nearly $50\%$. The cutoff $b$
must therefore be understood as a genuine physical parameter of the model
(the hard-sphere radius of paper I, representing the finite size of the
interacting particles), not as a regulator to be sent to zero at will.
\begin{table}[h]
\begin{tabular}{|ccc|}
\toprule
\hline\hline
$b\,(a_0)$ &\,& $E_{1,\ell=0}$ \\
\hline\hline
\midrule
0.05  &\,& $-0.09849$ \\
0.02  &\,& $-0.12379$ \\
0.01  &\,& $-0.13483$ \\
0.005 &\,& $-0.14111$ \\
0.001 &\,& $-0.14664$ \\
\hline\hline
\bottomrule
\end{tabular}
\centering
\caption{Ground-state energy ($\ell=0$) at $\kappa=0.5\,a_0^{-1}$,
$R=25\,a_0$, as a function of the inner cutoff $b$.}
\label{tab:bsens}
\end{table}

\textbf{Validity regime of the closed-form formulas : } The closed-form
expressions of Sections 3--4 use the \emph{free} hydrogenic basis as the
zeroth-order state, with energy $E_n^{(0)}=-\mu k^2/(2\hbar^2n^2)$, the
pure Coulomb value, uncorrected by the screening. Although the $\kappa$-dependence of $I(n,\ell;\kappa)$ is retained exactly in Eq. \ref{eq:Igeneral}, this does not make the full screened Yukawa spectrum analytic: the zeroth-order energy remains the Coulomb value. Table \ref{tab:validity} quantifies this mismatch
against the exact unconfined ground state ($b=0.05$, $R=60$): the deviation
grows from $16\%$ at $\kappa=0.01$ to $80\%$ at $\kappa=0.5$. The closed
forms should therefore be read as a useful analytic description of the
$(\Theta,\ell,\kappa)$ dependence of the NC correction, but not as a
quantitative prediction of the full spectrum outside $\kappa a_0\ll1$; all
final numbers in Section 6.4 come exclusively from the exact numerical
treatment of this section.
\begin{table}[h]
\begin{tabular}{|ccccccc|}
\toprule
\hline\hline
$\kappa\,(a_0^{-1})$ &\,& $E_{1,0}$ exact (unconfined) &\,& $E_1^{(0)}$ hydrogenic &\,& deviation \\
\hline\hline
\midrule
0.01 &\,& -0.42083 &\,& -0.5 &\,& -16\% \\
0.05 &\,& -0.38284 &\,& -0.5 &\,& -23\% \\
0.1  &\,& -0.33888 &\,& -0.5 &\,& -32\% \\
0.2  &\,& -0.26148 &\,& -0.5 &\,& -48\% \\
0.3  &\,& -0.19652 &\,& -0.5 &\,& -61\% \\
0.5  &\,& -0.09849 &\,& -0.5 &\,& -80\% \\
\hline\hline
\bottomrule
\end{tabular}
\centering
\caption{Ground-state energy: hydrogenic value (used in the closed forms)
versus exact unconfined value, as a function of $\kappa$.}
\label{tab:validity}
\end{table}

\subsection{Results}

\begin{figure}[h]
\centering
\includegraphics[width=0.75\textwidth]{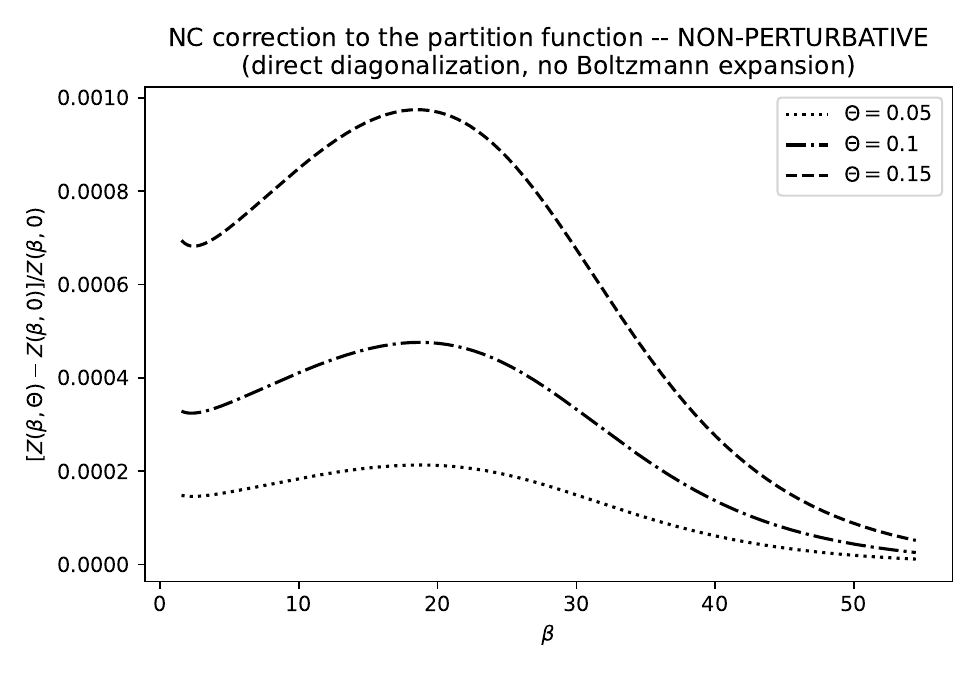}
\caption{NC correction (non-perturbative, direct diagonalization) to the
partition function, $[Z(\beta,\Theta)-Z(\beta,0)]/Z(\beta,0)$, converged
confined system, $\Theta=0.05,0.10,0.15$.}
\label{fig:Zfrac}
\end{figure}

\begin{figure}[h]
\centering
\includegraphics[width=0.75\textwidth]{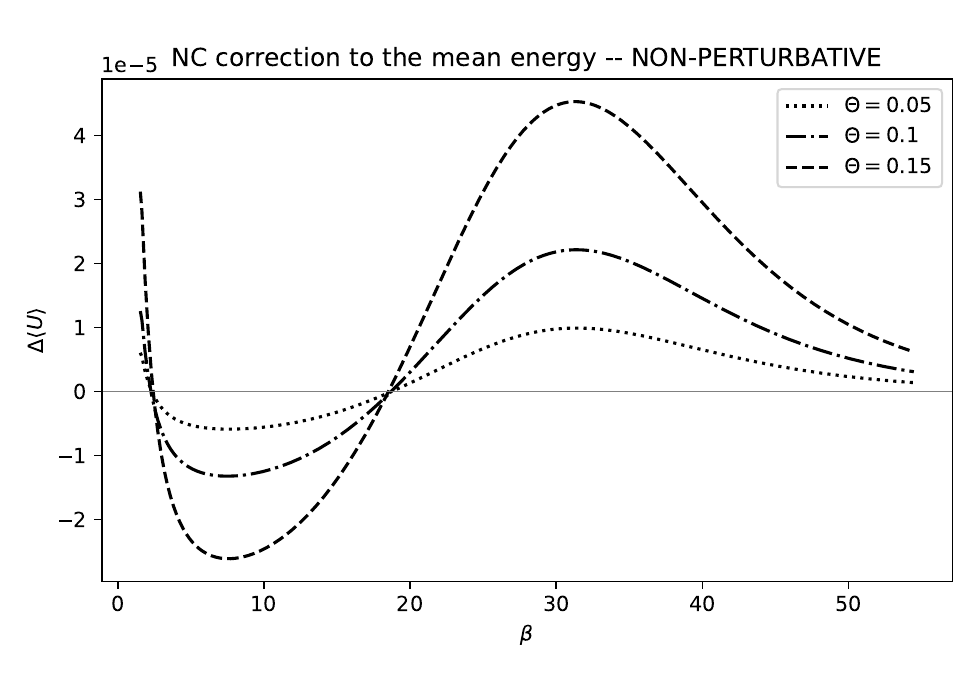}
\caption{NC correction (non-perturbative) to the mean energy,
$\Delta\langle U\rangle=\langle U\rangle(\beta,\Theta)-\langle
U\rangle(\beta,0)$. Note the three-zero-crossing structure, absent in the
linear approximation.}
\label{fig:dU}
\end{figure}

\begin{figure}[h]
\centering
\includegraphics[width=0.78\textwidth]{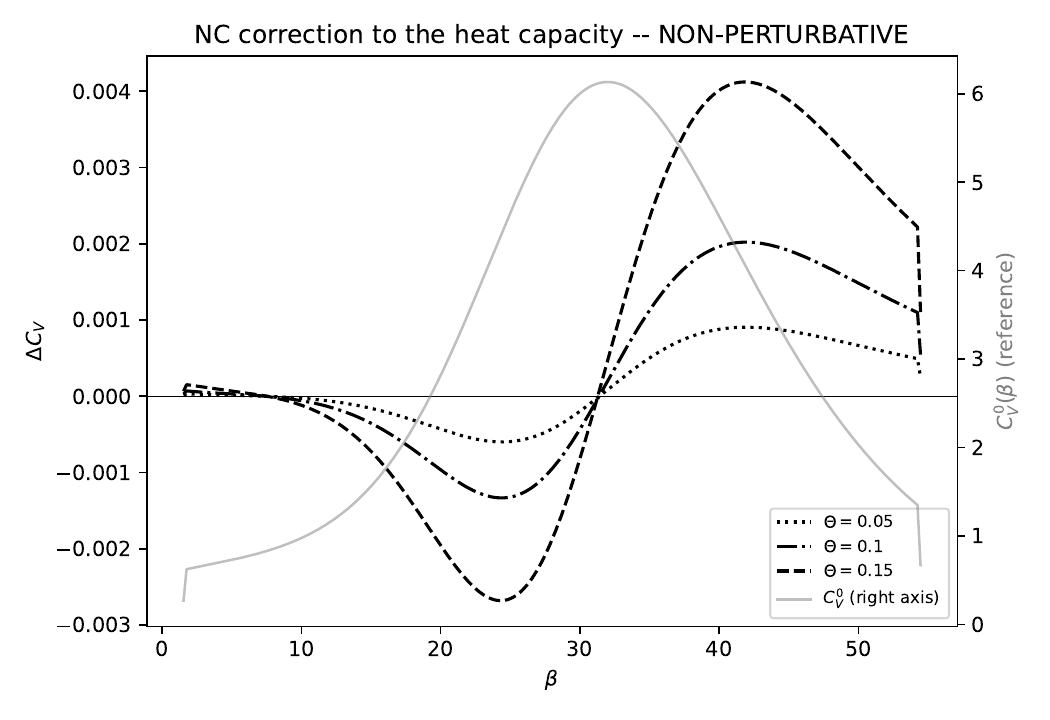}
\caption{NC correction (non-perturbative) to the heat capacity, $\Delta
C_V=C_V(\beta,\Theta)-C_V(\beta,0)$ (black lines, left axis), with
$C_V^0(\beta)$ reference (gray line, right axis). The negative dip
($\beta\approx24$) and the positive lobe ($\beta\approx41$) reach at most
$\approx0.07\%$ of the peak height, and the heat capacity remains positive
throughout. The plot range is trimmed to $\beta\in[1,52]$ to exclude
boundary artifacts of the numerical $\beta$-derivative.}
\label{fig:dCV}
\end{figure}

With the converged parameters and $\Theta=0.05,0.10,0.15$ all below the $\ell= 1$
threshold, we obtain the following results from the direct diagonalization described in Section \ref{sec:nonpert}.
The commutative baseline heat capacity has a Schottky peak near
$\beta\approx31.9$, with height approximately $6.13$. For $\Theta=0.15$,
the fractional partition-function correction shown in Fig. \ref{fig:Zfrac}
reaches about $0.098\%$ at $\beta\approx20$, roughly four times smaller
than the first-order prediction of about $0.4\%$ at the same point. The
mean-energy correction in Fig. \ref{fig:dU} has a richer structure than the
linear prediction, with sign changes near $\beta\approx2$--$3$ and
$18$--$19$ and extrema near $\beta\approx8$--$10$ and $30$--$32$. The heat
capacity correction in Fig. \ref{fig:dCV} likewise develops a negative dip
of approximately $-2.7\times10^{-3}$ near $\beta\approx24$ and a positive
lobe of approximately $4.1\times10^{-3}$ near $\beta\approx41$ for
$\Theta=0.15$. The largest deviation is about $0.07\%$ of the Schottky
peak height, while the total heat capacity remains positive throughout the
domain investigated.


\subsection{Comparison with the classical treatment}

For the Yukawa interaction, both the classical calculation of ref.\cite{Sousa2026} and the
present numerical quantum calculation yields non-negative heat capacity in the parameter ranges studied. The present results therefore show that the thermodynamic behavior found for
the Yukawa potential is qualitatively preserved when the radial Hamiltonian is
treated quantum mechanically.

\section{Phenomenological scale from the proton-neutron system}
\label{sec:pheno}

The Yukawa potential was originally introduced to describe a finite-range
interaction mediated by a massive exchanged particle \cite{Yukawa1935}.
In the nucleon-nucleon problem, one-pion exchange provides a standard
finite-range component of the interaction, although realistic deuteron
calculations require additional structure beyond a single central Yukawa
term \cite{Friar1984,Sprung1994}. This allows us to ask: using real
physical parameters of that system, what bound does the threshold
$\Theta_c^{\rm qu}$ of Section \ref{sec:fallcenter} place on the NC
parameter $\Theta$?

\subsection{Physical parameters}

The potential range is fixed by the pion mass,
$\kappa_{\rm phys}=m_\pi c/\hbar=138\,{\rm MeV}/(\hbar c)=0.699\,{\rm
fm}^{-1}$, and the reduced mass of the $p$-$n$ system is $\mu
c^2\approx469.5$ MeV. The bare one-pion-exchange coupling
($k=f^2\hbar c\approx 15.8$ MeV $\cdot$fm, with $f^2\approx0.08$) does not
bind the deuteron in this simplified central model,  physically expected,
since OPE alone is insufficient without intermediate meson-exchange
contributions. We therefore fit $k$ as an effective coupling to reproduce
the experimental deuteron binding energy, $E=-2.225$ MeV, solving
Eq.\eqref{eq:radial} numerically ($\ell=0$, $b=0.4$ fm): $k_{\rm
fit}=141.1$ MeV$\cdot$fm. This self-consistently defines
$a_0=\hbar^2/(\mu k_{\rm fit})=0.588$ fm and $\kappa_{\rm phys}a_0=0.411$, coincidentally close to the illustrative value $\kappa=0.5$ used
throughout the rest of this work \cite{Rosas}.

\subsection{Threshold in physical units}

Repeating the numerical search for $\Theta_c^{\rm qu}(\ell{=}1)$
(Section \ref{sec:fallcenter}) directly in physical units, with these
parameters and $b=0.4$ fm (typical nucleon-nucleon hard-core radius):
\begin{equation}
\Theta_c^{\rm qu}(\ell{=}1)\approx1.02 \, {\rm fm}^2,
\end{equation}
converged in grid resolution ($10^4$ to $10^5$ points). The sensitivity to
$b$ is significant (Table \ref{tab:thetac_b}):
\begin{table}[h]
\label{tab:thetac_b}
\begin{tabular}{|ccccc|}
\toprule
\hline\hline
$b$ (fm) &\,& $\Theta_c^{\rm qu}$ (fm$^2$) &\,& $\Theta_c^{\rm qu}/b$ \\
\hline\hline
\midrule
0.3 &\,& 0.662 &\,& 2.21 \\
0.4 &\,& 1.015 &\,& 2.54 \\
0.5 &\,& 1.481 &\,& 2.96 \\
0.6 &\,& 2.082 &\,& 3.47 \\
\hline\hline
\bottomrule
\end{tabular}
\centering
\caption{$\Theta_c^{\rm qu}(\ell{=}1)$ as a function of the nucleon-nucleon
hard-core radius.}
\end{table}

If $\Theta$ exceeds this scale, the cutoff-regularized model develops a deeply localized $P$-wave state. Interpreting the absence of such a state as a consistency requirement gives a model-dependent scale of order $\Theta\lesssim1$ fm$^2$; because it depends strongly on the assumed hard core, it is not a precision constraint.


The scale obtained here should not be identified directly with precision bounds
from atomic, Hall, or other systems. Such bounds depend on the precise
noncommutative deformation, the observable used, and the dimensional
normalization adopted. The present calculation instead provides a
model-dependent nuclear consistency scale associated with the appearance of
a cutoff-localized state. A quantitative comparison with other constraints
would require rederiving each bound in exactly the same deformation convention.


The main physical content of the phenomenological exercise is therefore not a
competitive experimental limit. Rather, it demonstrates that the quantum
collapse mechanism can be translated into a nuclear length scale and that the
corresponding threshold is strongly controlled by the assumed short-distance
core. Within the present effective model, the absence of a deeply localized
collapse state would imply the consistency condition
$\Theta\lesssim\Theta_c^{\rm qu}$, with the numerical value determined
by the chosen interaction and cutoff.

\section{Conclusions and outlook}

We have studied a quantum realization of the noncommutative Yukawa
interaction obtained by quantizing the radial Hamiltonian obtained in the ref. \cite{Sousa2026}.
We adopted the prescription of $M\to\hbar \, \sqrt{\ell(\ell+1)}$ in the
already-reduced radial Hamiltonian. Within this prescription the spectrum
depends on $(n,\ell)$ but not on $m_\ell$-quantum numbers, preserving the $(2\ell+1)$-
degeneracy. The first-order NC energy shift can be written in closed form
for the hydrogen reference states: the screening parameter $\kappa$ is
retained exactly in the perturbation matrix element, while the reference
energy remains Coulombian. These formulas are therefore analytic expressions
for the NC correction rather than an exact analytic solution of the screened
Yukawa spectrum.

The classical and quantum instability criteria are distinct. In the Coulomb
limit the classical circular orbit disappears at
$\Theta_c^{\rm cl}=[\ell(\ell+1)]^{3/2}/12$, whereas the quantum calculation
reveals a cutoff-dependent onset of a state localized near the inner boundary.
For $b=0.05a_0$ and $\kappa=0.5a_0^{-1}$, the numerical thresholds are
$\Theta_c^{\rm qu}=0.176$, $0.202$, and $0.229$ for $\ell=1,2,3$,
respectively. Above these values the lowest state becomes increasingly
localized near the inner boundary, and its energy grows in magnitude as the
cutoff is reduced. The threshold
increases with $b$ over the examined range; this trend motivates a systematic study $\Theta_c^{\rm qu}(\ell,b,\kappa)$.

The thermodynamic quantities were evaluated from the full cutoff-regularized
spectrum by direct diagonalization, without expanding the Boltzmann factor.
For $\Theta=0.15$, the fractional correction to the partition function is about $0.098\%$ at $\beta=20$, compared with about $0.407\%$ from the first-order estimate at the same point. Thus first order overestimates the
thermodynamic correction by roughly a factor of four in this representative case. The finite-system identity
$C_V=\beta^2(\langle E^2\rangle-\langle E\rangle^2)\ge0$ ensures
non-negative heat capacity, and the numerical results show that the NC interaction shifts and reshapes the Schottky structure without producing a negative heat-capacity region in the investigated parameter range.

Finally, the simplified proton-neutron model with $b=0.4$ fm gives
$\Theta_c^{\rm qu}(\ell=1)\approx1.02$ fm$^2$. Because this number depends substantially on the hard-core radius and on the effective interaction model, it is best interpreted as a model-dependent short-distance consistency scale rather than as a precision experimental bound. Natural extensions include a
systematic map of $\Theta_c^{\rm qu}(\ell,b,\kappa)$, a semiclassical/WKB analysis connecting the quantum threshold with the classical effective potential, an explicit second-order perturbative calculation, and application
of the same procedure to the noncommutative Lee--Wick potential. A
Beth--Uhlenbeck or Planck--Larkin treatment could also provide an alternative thermodynamic formulation without relying exclusively on the auxiliary outer box. This idea can be a way for future investigations.

\end{document}